# High-power few-cycle near-infrared OPCPA for soft X-ray generation at 100 kHz

Stefan Hrisafov*, Justinas Pupeikis, Pierre-Alexis Chevreuil, Fabian Brunner, Christopher R. Phillips, Lukas Gallmann, and Ursula Keller

Department of Physics, Institute for Quantum Electronics, ETH Zurich, Auguste-Piccard-Hof 1, 8093 Zurich, Switzerland

*Corresponding author: hrisafov@phys.ethz.ch

**Abstract**

We present a near-infrared optical parametric chirped-pulse amplifier (OPCPA) and soft X-ray (SXR) high-harmonic generation system. The OPCPA produces few-cycle pulses at a center wavelength of 800 nm and operates at a high repetition rate of 100 kHz. It is seeded by fully programmable amplitude and phase controlled ultra-broadband pulses from a Ti:sapphire oscillator. The output from the OPCPA system was compressed to near-transform-limited 9.3-fs pulses. High-power operation up to an average power of 35 W was achieved, and a fully characterized pulse compression was recorded for a power level of 22.5 W, demonstrating pulses with a peak power greater than 21 GW. We demonstrate that at such high repetition rates, spatiotemporally flattened pump pulses can be achieved through a cascaded second-harmonic generation approach with an efficiency of more than 70%, providing a compelling OPCPA architecture for power-scaling ultra-broadband systems in the near-infrared. The output of this 800-nm OPCPA system was used to generate SXR radiation reaching 190 eV photon energy through high-harmonic generation in helium.

## 1. Introduction

The recent progress of multi-100-Watt and high-repetition-rate picosecond laser oscillators and amplifier systems at 1.03 μm [1-4], enabled the development of few-cycle driving sources for attosecond science at repetition rates ranging from 100 kHz to megahertz [5-7]. This resulted in the first high-repetition-rate high-harmonic generation (HHG) sources extending into the soft X-ray (SXR) spectral range (>124 eV) [8,9] and the water-window SXR radiation [5,10]. The higher pulse repetition rate allows for higher photon flux in the HHG output, leading to improved signal-to-noise ratio (SNR) and reduced acquisition times in attosecond experiments [11,12]. For example, for photoemission, space-charge effects can be reduced [13], and for coincidence detection, especially, better statistics or shorter acquisition times are essential [14,15].

Optical parametric chirped-pulse amplification (OPCPA) [16,17] currently represents one of the most compelling approaches for combining the high average powers of the high-repetition-rate picosecond pump lasers with the ultrabroad bandwidth of Ti:sapphire lasers or supercontinuum-generated seed pulses. The compressed few-cycle output can yield a high-repetition-rate source with the necessary peak intensity for HHG and strong-field experiments. Closely tied to the progress in pump laser technology, power- and repetition-rate-scaling have therefore strongly driven the development of near-infrared (near-IR) OPCPA systems in recent years [7,18,19].

One of the main challenges in the power-scaling of near-IR OPCPA is the efficient generation of the pump pulses. Since optical parametric amplification (OPA) around 800 nm requires a pump in the visible spectral range, the ultimate efficiency of a near-IR OPCPA system is heavily dependent on the efficient second-harmonic generation (SHG) of the 1–µm pump pulses. The highest experimentally-observed SHG efficiencies for picosecond high-repetition-rate lasers in the multi-hundred-Watt average power range are limited to 70% [20,21] and have been achieved with a single-stage SHG in Lithium Triborate (LBO) or beta-Barium Borate (BBO). A very promising approach for improving the SHG efficiency is cascaded SHG [22] which has been demonstrated with efficiencies in the range of 60-70% at 1-kHz repetition rate [18,22], and very recently with 85% for a 0.1-kHz system [23]. The first stage in cascaded SHG is used to deplete the most intense parts of the fundamental pulses, resulting in a spatiotemporally flattened fundamental at the output. These flattened fundamental pulses are then re-used to drive a second SHG stage. Such a flattened pump pulse profile can increase the efficiency of high-power OPA stages, as it lifts the saturation limit to higher pulse energies and reduces the detrimental effects of spectral gain narrowing which is a major concern for efficient few-cycle OPCPA systems [23,24].

Here, we present a high-power 100-kHz near-IR OPCPA system delivering an average power of 22.5 W (225 µJ) at a center wavelength of 800 nm with a compressed pulse duration of 9.3 fs (3.5 cycles) and a peak power of 21 GW. High-power OPCPA operation was also observed at the output power level of 35 W without compression. To achieve this OPCPA performance, we leverage the cascaded SHG approach to generate the high-power and high-repetition-rate pump beams that drive the final OPCPA amplification stages. To demonstrate the practical use of this source, we present first experimental results of HHG in a high-pressure helium target generating SXR radiation reaching photon energies of 190 eV.

## 2. Experimental set-up of the OPCPA

The full system is schematically depicted in Fig. 1. The OPCPA chain consists of 4 non-collinear OPA stages in beta-barium borate (BBO), and can be split in two main sections – a pre-amplification section, which includes two OPA stages (OPA1 and OPA2) and their respective pump generation (SHG1), as well as a power-amplification section, with OPA3 and OPA4 and a high-power cascaded SHG set-up (SHG2 and SHG3).

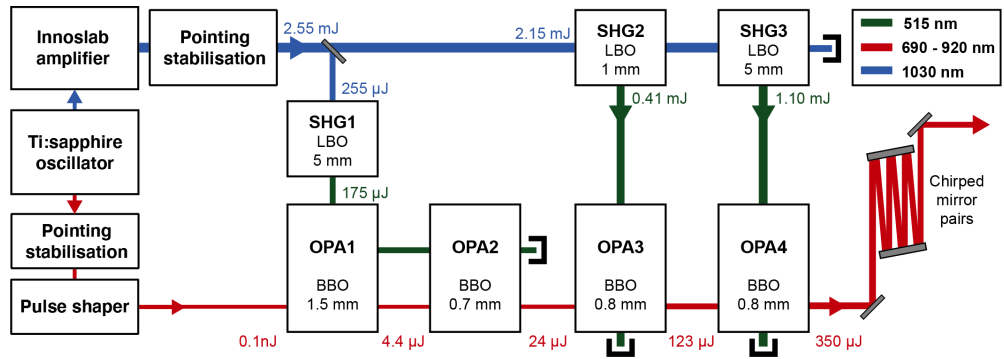

Fig. 1. The full OPCPA set-up. BBO: beta-barium borate. LBO: Lithium triborate. OPA: Optical Parametric Amplifier. SHG: Second Harmonic Generation.

We use a Ti:sapphire oscillator (Venteon PulseONE; Laser Quantum GmbH) to seed both the OPCPA chain as well as the pump laser - an Innoslab-based laser amplifier system (A400; Amphos GmbH). The seed laser produces close to transform-limited pulses with sub-6-fs duration at a repetition rate of 82 MHz with an average power of 210 mW. We actively lock the carrier-envelope offset frequency [25] to a quarter of the repetition rate of the laser. A selected portion of its spectrum around 1030 nm is used to seed the pump amplifier chain after pre-amplification and pulse-picking down to the final 100 kHz repetition rate. The pulse picking is chosen to be a multiple of 4, such that the OPCPA system will only amplify pulses from the 82 MHz pulse train that have the same carrier-envelope-phase (CEP). This ensures a passive CEP-stability of the OPCPA output, with an active stabilization of any slow drifts yet to be implemented. The rest of the broadband seed spectrum goes through a 4-f pulse-shaper based on a high-resolution spatial-light-modulator (SLM). The output from this pulse shaper spans a spectral bandwidth from 640 nm to 920 nm and is used to seed the OPA1. The 4-f scheme and the SLM, together with OPA1 are used as a time-gated pulse-shaper, which we have previously reported in [26]. This powerful pulse-shaping scheme allows for both a coarse and fine programmable control of the spectral phase and amplitude of our seed pulses yielding a flexible OPCPA output for experiments. The pump laser provides 2-ps-long pulses at a center wavelength of 1030 nm and an average power of 255 W (corresponding to a pulse energy of 2.55 mJ). The beam-pointing of both the pump and seed laser outputs is actively stabilized with commercial stabilization units (TEM Messtechnik GmbH) in order to ensure stable pump and seed overlap in the OPCPA chain.

### 3. Pump shaping through cascaded second harmonic generation

For the SHG we use low-absorption LBO crystals ($\theta$=90°, $\phi$=12.8°, type I; Cristal-Laser SA) in temperature-controlled mounts (stabilized at room temperature, i.e. 21°C, in order to avoid any thermal load in the set-up while still maintaining a stable temperature). A small fraction (10%) of the available pump light at 1030 nm is used to generate the pump for the OPCPA pre-amplifiers in the SHG stage labelled SHG1 in Fig. 1. We use a 5-mm-long LBO crystal to generate 17.5 W of second harmonic light with an efficiency of 69%.

Of the remaining fundamental pump light at 1030 nm, 215 W reach the high-power SHG set-up, which consists of two consecutive stages with 1-mm and 5-mm-long LBO crystals, respectively (SHG2 and SHG3 in Fig. 1). We employ a cascaded SHG approach due to its ability to produce spatiotemporally flattened pump pulses for the last amplification stage. This choice is motivated with Fig. 2, which shows a simulation of the pulse profiles in each stage of the cascaded SHG, following the approach for a phase-matched SHG discussed in [27].

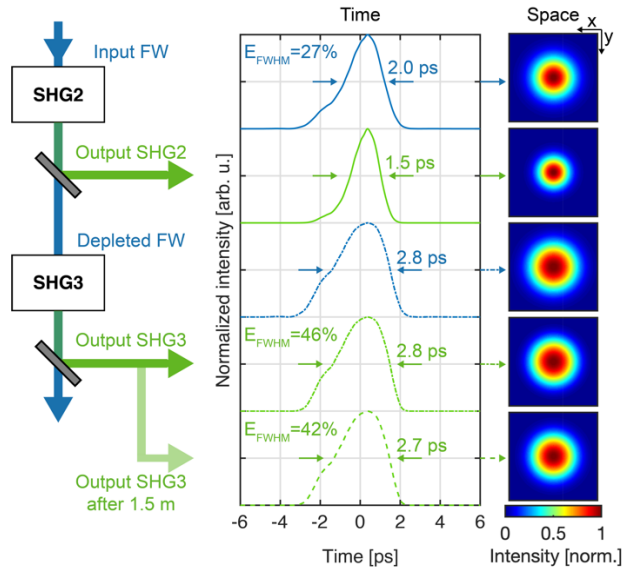

Fig. 2. Simulated on-axis temporal and time-averaged spatial profiles (assuming perfect phase-matching) of the fundamental wave (FW) in the high-power cascaded 2-stage SHG (labelled as SHG2 and SHG3), and the subsequent pulse profiles of the second harmonic from both stages. The designed SHG2 efficiency is 23%. The pulse profile of the SHG3 output after 1.5 m was calculated according to Rayleigh-Sommerfeld diffraction. $E_{FWHM}$ represents the energy fraction contained within the full-width-at-half-maximum of the pulse profile

The simulation assumes a Gaussian spatial profile for the fundamental input that was matched to the experimentally measured beam size. The temporal profile used in the simulation was experimentally measured with an SHG-FROG in a 0.3-mm-long BBO-crystal. Fig. 2 depicts the resulting on-axis temporal profiles of the fundamental and the second harmonic, as well as the corresponding time-averaged spatial profiles, throughout both high-power SHG stages. The fraction of the energy contained within the full-width-at-half-maximum (FWHM) of the pulse profile represents a good metric for judging the quality of the pump source. This is because the small-signal gain of the OPA process scales nonlinearly with the pump peak intensity. Therefore, the design of the cascaded SHG was optimized to increase this fraction in the pump pulses for the last OPA stage, through spatiotemporal flattening of the fundamental pulses in SHG2 while at the same time maintaining sufficient pulse energy in the second harmonic output from SHG3. As such, the chosen design allows favorable pump energy ratio between the two final OPA stages, while also providing relatively sufficient spatiotemporal flattening in the cascaded SHG.

The fundamental input pulses have a pulse duration of 2 ps, with 27% of the pulse energy contained within the FWHM (assuming a Gaussian beam profile). The first stage of the cascaded SHG, i.e. SHG2, is designed for an efficiency of 23%, which should result in a depleted fundamental with an on-axis pulse duration of 2.8 ps. The spatiotemporal flattening is almost fully transferred to the second harmonic in SHG3, with the same estimated pulse duration of 2.8 ps, resulting in 46% of the pulse energy within the FWHM. The total designed SHG efficiency (both stages together) is 87%. Finally, in Fig. 2, we also present the pulse profile of the second harmonic output from SHG3 propagated after 1.5 m (simulated according to Rayleigh-Sommerfeld diffraction), which corresponds to the position of the final OPA stage, OPA4. Due to this propagation, the spatial profile slowly reverts towards a Gaussian profile, resulting in an expected 42% energy within the FWHM available to pump OPA4. This translates to an effective increase in the energy within the FWHM by 56%, demonstrating the utility of this flattening scheme.

In practice, the output power from SHG2 was 41 W at 515 nm, while the output power from SHG3 was 110 W, resulting in a total efficiency of the high-power SHG of 70%. When calculating the efficiency, the losses from various optics between both stages (~3%) have not been compensated for; correcting for these, we estimate a conversion efficiency of 73%. The discrepancy with respect to the simulated efficiency is not yet understood. It is possible that unwanted additional phase shifts due to third-order nonlinear effects or thermal lensing effects could cause a partial dephasing of the SHG process. Another possibility is light from potential amplified spontaneous emission co-propagating with the output pulses of the laser amplifier which would not contribute to the SHG.

A summary of the experimental parameters of all SHG stages is shown in Table 1. The peak intensities presented for SHG1 and SHG2 use the experimentally characterized pulse duration of the laser output, while the intensity for SHG3 uses the estimated pulse duration of the depleted fundamental as shown in Fig. 2 with a measured depleted beam shape.

Table 1. Summary of the performance of all second harmonic stages in the system

|  | LBO length [mm] | $P_{1030}$ [W] | Peak intensity [GW/cm$^2$] | $P_{515}$ [W] | Efficiency [%] |
|---|---|---|---|---|---|
| **SHG1** | 5 | 25.5 | 24 | 17.5 | 69 |
| **SHG2** | 1 | 215 | 19 | 41 | 19 |
| **SHG3** | 5 | 174 | 10 | 110 | 63 |
| **Total SHG2+3** | (1 and 5) | 215 | - | 151 | 70 |

## 4. Pre-amplifiers and power-amplifiers – experimental details

The pre-amplifiers OPA1 and OPA2 are pumped with the 17.5 W of second harmonic generated in SHG1. Both are in the non-collinear type-1 interaction geometry, with an internal non-collinear angle of ~2.45°. This angle minimizes the group velocity mismatch between the signal and the idler and thus provides the broadest amplification bandwidth. Additionally, because the walk-off compensating configuration leads to phase-matched SHG of the signal around 870 nm, we use the alternative non-walk-off compensating configuration [28] in all of the OPA stages. The first OPA stage uses a 1.5-mm-long BBO crystal ($\theta$=23°, $\phi$=90°; EKSMA Optics) pumped at a peak intensity of 130 GW/cm$^2$ (i.e. 1/e$^2$ beam radius of ~0.2 mm), resulting in a gain of 46 dB (Table 2). In the calculation of the gain and efficiency in Table 2, the reflection losses from the crystal facets, which are coated only with a protective coating, were not compensated for.

The seed and the amplified spectra from each OPA stage are shown in Fig. 3(a). The output spectrum of OPA1 spans the bandwidth from 690 nm to 920 nm and is maintained throughout the amplification chain. The seed spectrum between 640 nm and 690 nm is intentionally not amplified in favor of the longer-wavelength part of the spectrum. The leftover pump from OPA1 is re-used for OPA2, which uses a 0.7-mm-long BBO crystal ($\theta=24°$, $\phi=90°$; EKSMA Optics), producing 2.4 W of average power in the signal. Both the pump and signal are imaged from OPA1 to OPA2 to ensure reliable and stable operation of the pre-amplifiers.

To account for long-term temporal drifts between the pump and seed, time-delay stabilization has been implemented within OPA1, using the center wavelength of the signal spectrum from OPA1 as the reference for maintaining the temporal overlap. This is achieved with a feedback loop that acts on a motorized stage placed in the seed beam path.

Table 2. Summary of the performance of the four OPA stages in the OPCPA chain

|       | BBO length [mm] | Pump peak intensity [GW/cm2] | Signal output energy [µJ] | Gain [dB] | Efficiency [%] |
|-------|-----------------|------------------------------|---------------------------|-----------|----------------|
| OPA1  | 1.5             | 130                          | 4.4                       | 46        | 2.5            |
| OPA2  | 0.7             | 120                          | 24                        | 6.5       | 11.2           |
| OPA3  | 0.8             | 60                           | 123                       | 6.2       | 24.2           |
| OPA4  | 0.8             | 30                           | 350                       | 2.7       | 20.6           |

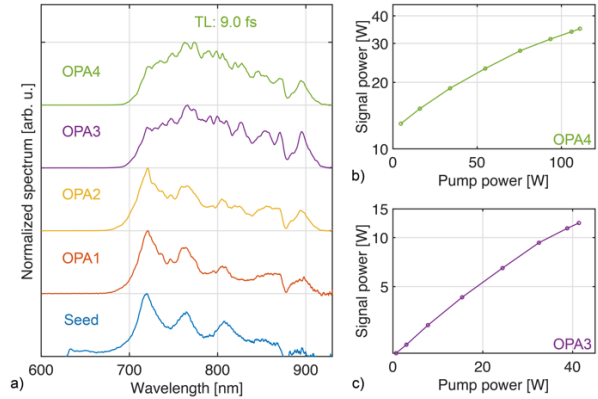

Fig. 3. (a) Normalized optical spectrum of each OPA stage output. TL: transform-limited pulse duration calculated from the output spectrum, λc: center wavelength of the final output spectrum. Power slope measurements of the power amplifiers (b) OPA4 and (c) OPA3.

The power-amplifiers, OPA3 and OPA4, are both based on 0.8-mm-long BBO crystals ($\theta=23°$, $\phi=90°$; EKSMA Optics) operated in the same non-walk-off-compensating non-collinear geometry for type-I nonlinear interaction. The pump peak intensities in OPA3 and OPA4 are 60 GW/cm$^2$ and 30 GW/cm$^2$, respectively, estimated from the temporal profiles in Fig. 2 and using an experimentally measured beam shape. The output spectrum of OPA3, shown in Fig. 3(a), suggests a slight gain narrowing, due to the shorter pump pulse duration from the non-saturated SHG2. This has been pre-compensated in the preamplifiers such that the final spectrum more closely resembles a flat-top spectrum (ensuring a short Fourier-transform limit). OPA4 then maintains the same spectrum, owing to its significantly flatter pump profile compared to OPA3. Additionally, both power-amplifiers are operated within the saturation regime, but not yet at the limit of back-conversion as is demonstrated in the measured power slopes in Fig. 3(b-c), for a final average power of 35 W.

Due to its inverse proportionality to wavelength, the contributions to the accumulated B-integral from any transmissive material, including air, are much stronger at 515 nm compared to 1030 nm. Therefore, to maintain a stable and clean spatiotemporal pump profile for both power amplifiers, we perform pump shaping with curved dielectric mirrors only. Additionally, the pump light is not imaged from the high-power cascaded SHG stages to the OPA crystals. This is because such imaging would require a tight focus within the beam path, leading to excessive self-phase modulation from the nonlinear refractive index of air. Due to the large beam sizes in the cascaded SHG, the effect of propagation to the spatiotemporal profile is weak enough to allow for sufficiently flat pump pulses in OPA4, as shown with the calculated pulse profiles in Fig. 2. Contrary to the pump, the signal is shaped in the power amplifiers with UV-grade fused silica (UVFS) lenses, thanks to its lower power as well as the lower B-integral at that wavelength. Additionally, the transmissive shaping optics contribute to the stretching of the signal pulses between the OPA stages.

## 5. Dispersion management and compression

Most of the stretching of the near-IR pulses comes from the positive group-delay dispersion (GDD) accumulated from the shaping optics, which include N-BK7 lenses in the preamplifiers, and UVFS lenses in the power amplifiers. Additional stretching comes from flat N-BK7 and SF11 windows. The relatively large positive third-order dispersion (TOD) that is accumulated throughout the amplification chain is pre-compensated at the beginning of the system using our time-gated pulse-shaping scheme [26] with minimal spectral aberrations. The pulse shaper is additionally used for fine dispersion compensation during compression. As can be seen in Fig. 4(a), where we show the accumulated group delay for each stage in the OPCPA chain, we apply a large negative TOD ($\sim -5000$ fs$^3$) on the seed pulses, which results in an almost fully GDD-dominated phase in OPA4 that allows for easy compression and clean pulses.

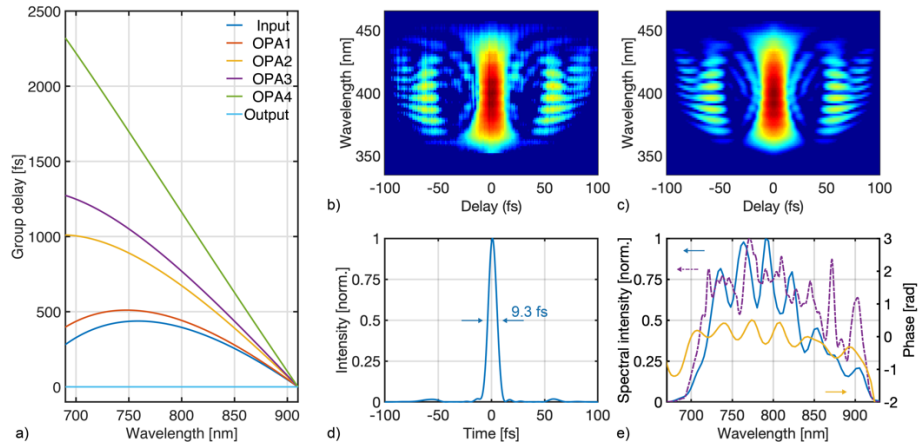

Fig. 4. (a) Group delay accumulated for each stage in the OPCPA chain. (b) Measured and (c) reconstructed FROG trace of the compressed output pulse, with a reconstruction error of 0.5% using a grid of 256 (in time) x 416 (frequency) points. (d) Reconstructed temporal profile. (e) The reconstructed spectrum (blue full line) and spectral phase (yellow full line), overlaid with the measured spectrum (purple dashed line). Before reconstruction, the FROG trace was corrected for the phase-matching in the SHG and the non-collinear geometry according to [29].

We first performed a full characterization of the OPCPA output at an intermediate average power of 22.5 W. The compression of the OPCPA output is done with 32 reflections on chirped mirror pairs (Layertec GmbH), resulting in a total GDD of -3800 fs$^2$ and a TOD of 2600 fs$^3$ at 800 nm. The compressor throughput is around 90%, yielding the reported 22.5 W after compression. The compressed pulses have been characterized with SHG-FROG in a 20-µm-long BBO crystal. The corresponding FROG trace, reconstructed pulse shape, as well as the final output spectrum are shown in Fig. 4(b-e). The compressed FWHM pulse duration of 9.3 fs corresponds to 3.5 cycles at the center wavelength of 800 nm (estimated from the center of gravity of the experimentally measured spectrum from Fig. 4(e)). From the pulse shape and energy, we determine a peak power of 21 GW. The remaining ripples in the spectral phase originate from residual dispersion oscillations of the chirped-mirrors in the compressor and can be removed with an optimized feedback to the programmable pulse shaper.

After the above-mentioned characterization, we further optimized the efficiency of the OPCPA, through improved stretching and spatial shaping in the power amplifiers, resulting in a higher output power of 35 W directly after OPA4, with the same ultra-broadband output. This 35 W configuration is the one described in section 4, and based on the compressor throughput we would expect >30 W after compression. However, the performance of the pump laser degraded during the course of this optimization, preventing complete temporal characterization. Hence, we have demonstrated compressed OPCPA output at 22.5 W and a power-optimized but uncompressed OPCPA at 35 W.

## 6. High-harmonic generation in helium

We use 18 W (180 µJ) of the near-IR OPCPA output for HHG in helium using the same HHG module as for our recently reported demonstration of water-window HHG from a mid-IR OPCPA [5]. The generated high harmonics are characterized with a CCD-based (Newton 400; Andor) flat-field grating-spectrometer (251MX; McPherson). At the entrance of the HHG chamber, we use an anti-reflection (AR) coated UVFS window and a plano-convex AR-coated UVFS lens with a focal length of 75 mm. This led to an experimentally measured $1/e^2$ beam diameter of 33 µm at the focus. The dispersion from the focusing lens and the entrance window was taken into account during compression. The helium gas was supplied from a needle with an inner diameter of 0.4 mm and 0.1-mm-thick walls. The stainless-steel needle was drilled by the near-IR beam, ensuring a relatively constant interaction length at stable pressure.

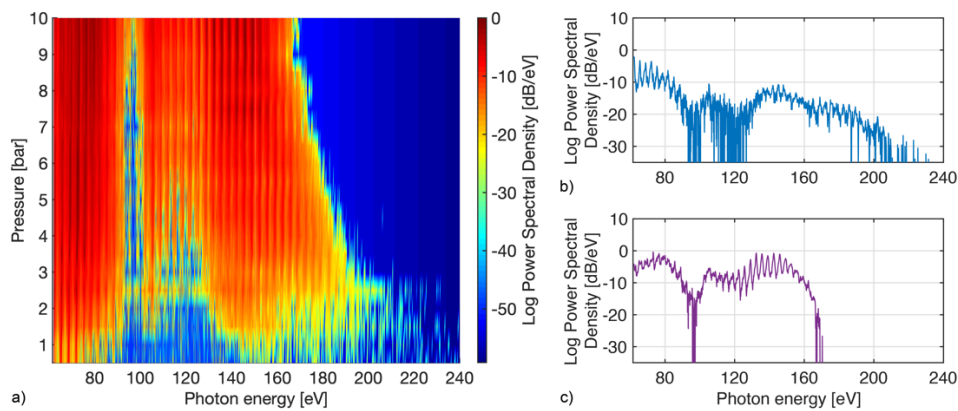

Fig. 5. (a) Pressure dependence of the spectrum from HHG in helium. Selected spectra at pressures of 2.5 bar (b) and 10 bar (c).

In Fig. 5(a), we present the spectrum of the generated XUV/SXR radiation as a function of helium backing pressure, measured for a pressure range between 0.5 bar and 10 bar in steps of 0.5 bar. All presented spectra were recorded for a spectrometer entrance slit width of 1 mm and with a 100-nm-thick aluminium (Al) filter for blocking any leftover near-IR light. The spectrometer exposure time was 1 s. The amplitude of each displayed spectrum is corrected for the CCD response, the grating reflectivity, as well as the filter transmission. The generated radiation was optimized for flux at the highest cut-off for the pressure of 2.5 bar, by translating the position of the beam focus along the propagation direction. This resulted in the spectrum shown in Fig. 5(b), with a cut-off at 190 eV. From this cut-off we can estimate an achieved peak intensity of 870 TW/cm$^2$. We expect a plasma defocusing effect to be the limiting factor in the obtainable photon energies in the current geometry. Due to lack of a calibrated measurement device, the absolute flux was not measured. Instead, we estimate the lower boundary of the flux from the CCD counts for the spectral range between 62 eV and 248 eV, as covered by the grating used in the spectrometer (1200 grooves/mm). Any potential clipping at the spectrometer entrance slit is not taken into account. The spectrum shown in Fig. 5(b) therefore leads to an estimated flux of 0.2 nW. For comparison, in Fig. 5(c), we show the spectrum for a backing pressure of 10 bar, which resulted in photon energies up to 160 eV, but with a higher total flux, estimated as 1.1 nW for the spectral range between 62 eV and 248 eV.

## 7. Conclusion

We presented a high-power near-IR few-cycle OPCPA system generating nearly-transform-limited pulses with a duration of 9.3 fs. The pulses are centered at a wavelength of 800 nm and carry an energy of 225 µJ (22.5 W). This translates to a peak power of more than 21 GW. Furthermore, we report operation of the system at energies up to 350 µJ (35 W) without compression. We demonstrate the effectiveness of the cascaded SHG process for the generation of high-power and high-repetition-rate second harmonic from the 1030-nm pump laser, with more than 70% of efficiency, as a compelling approach for power-scaling near-IR OPCPA systems. Further development of the system will include characterizing the compression when operating at 35 W of average power, as well as stabilizing the slow drifts of the CEP.

The few-cycle near-IR OPCPA was used for driving HHG in helium, generating XUV and SXR radiation at photon energies up to 190 eV. This result represents one of the first steps in the scaling of the obtainable photon energies from a high-repetition-rate 800-nm-based HHG source. Future work will include optimizing the driving peak intensity by improving the focusing and interaction geometry, with the potential goal to provide a source of high-flux SXR radiation at 100 kHz and thus an alternative to the mid-infrared OPCPA systems currently used for the generation of high-photon-energy harmonics.

**Funding**

This work was supported by the Swiss National Science Foundation (SNSF) projects 200020_172644, and 206021_164034/1.

**References**

1. P. Russbueldt, T. Mans, J. Weitenberg, H. D. Hoffmann, and R. Poprawe, "Compact diode-pumped 1.1 kW Yb:YAG Innoslab femtosecond amplifier," Opt. Lett. **35**, 4169–4171 (2010)
2. T. Eidam, S. Hanf, E. Seise, T. V. Andersen, T. Gabler, C. Wirth, T. Schreiber, J. Limpert, and A. Tünnermann, "Femtosecond fiber CPA system emitting 830 W average output power," Opt. Lett. **35**, 94–96 (2010)
3. J.-P. Negel, A. Loescher, A. Voss, D. Bauer, D. Sutter, A. Killi, M. A. Ahmed, and T. Graf, "Ultrafast thin-disk multipass laser amplifier delivering 1.4 kW (4.7 mJ, 1030 nm) average power converted to 820 W at 515 nm and 234 W at 343 nm," Opt. Express **23**, 21064–21077 (2015)
4. F. Saltarelli, I. J. Graumann, L. Lang, D. Bauer, C. R. Phillips, and U. Keller, "Power scaling of ultrafast oscillators: 350-W average-power sub-picosecond thin-disk laser," Opt. Express **27**, 31465–31474 (2019)
5. J. Pupeikis, P.-A. Chevreuil, N. Bigler, L. Gallmann, C. R. Phillips, U. Keller, "Water window soft x-ray source enabled by a 25 W few-cycle 2.2 µm OPCPA at 100 kHz", Optica **7**, 168-171 (2020)
6. T. Nagy, S. Hädrich, P. Simon, A. Blumenstein, N. Walther, R. Klas, J. Buldt, H. Stark, S. Breitkopf, P. Jójárt, I. Seres, Z. Várallyay, T. Eidam, and J. Limpert, "Generation of three-cycle multi-millijoule laser pulses at 318 W average power", Optica **6**, 1423-1424 (2019)
7. F. J. Furch, T. Witting, A. Giree, C. Luan, F. Schell, G. Arisholm, C. P. Schulz, and M. J. J. Vrakking, "CEP-stable few-cycle pulses with more than 190 µJ of energy at 100 kHz from a noncollinear optical parametric amplifier," Opt. Lett. **42**, 2495-2498 (2017)
8. J. Rothhardt, S. Hädrich, A. Klenke, S. Demmler, A. Hoffmann, T. Gotschall, T. Eidam, M. Krebs, J. Limpert, and A. Tünnermann, "53 W average power few-cycle fiber laser system generating soft x rays up to the water window," Opt. Lett. **39**, 5224-5227 (2014)
9. R. Klas, W. Eschen, A. Kirsche, J. Rothhardt, and J. Limpert, "Generation of coherent broadband high photon flux continua in the XUV with a sub-two-cycle fiber laser," Opt. Express, **28**, 6188-6196 (2020)

10. M. Gebhardt, T. Heuermann, Z. Wang, M. Lenski, C. Gaida, R. Klas, A. Kirsche, S. Hädrich, J. Rothhardt, and J. Limpert, "Soft x-ray high order harmonic generation driven by high repetition rate ultrafast thulium-doped fiber lasers, " Proc. SPIE 11260, Fiber Lasers XVII: Technology and Systems, 112600U (2020)

11. F. Krausz and M. Ivanov, "Attosecond physics," Rev. Mod. Phys. **81**, 163–234 (2009)

12. L. Gallmann, C. Cirelli, and U. Keller, "Attosecond Science: Recent Highlights and Future Trends," Annu. Rev. Phys. Chem. **63**, 447–469 (2012)

13. S. Passlack, S. Mathias, O. Andreyev, D. Mittnacht, M. Aeschlimann, and M. Bauer, "Space charge effects in photoemission with a low repetition, high intensity femtosecond laser source," J. Appl. Phys. **100**, 024912 (2006)

14. J. Ullrich, R. Moshammer, A. Dorn, R. Dorner, L. P. H. Schmidt, and H. Schmidt-Bocking, "Recoil-ion and electron momentum spectroscopy: reaction-microscopes," Rep. Prog. Phys. **66**, 1463–1545 (2003)

15. M. Sabbar, S. Heuser, R. Boge, M. Lucchini, L. Gallmann, C. Cirelli, and U. Keller, "Combining attosecond XUV pulses with coincidence spectroscopy," Rev. Sci. Instrum. **85**, 103113 (2014)

16. A. Dubietis, G. Jonušauskas, and A. Piskarskas, "Powerful femtosecond pulse generation by chirped and stretched pulse parametric amplification in BBO crystal", Opt. Commun. **88**, 437-440 (1992)

17. S. Witte and K. S. E. Eikema, "Ultrafast Optical Parametric Chirped-Pulse Amplification," IEEE Journal of Selected Topics in Quantum Electronics **18**, 296-307 (2012)

18. R. Budriūnas, T. Stanislauskas, J. Adamonis, A. Aleknavičius, G. Veitas, D. Gadonas, S. Balickas, A. Michailovas, and A. Varanavičius, "53 W average power CEP-stabilized OPCPA system delivering 5.5 TW few cycle pulses at 1 kHz repetition rate", Opt. Express, **25**, 5797-5806 (2017)

19. K. Mecseki, M. K. R. Windeler, A. Miahnahri, J. S. Robinson, J. M. Fraser, A. R. Fry, and F. Tavella, "High average power 88 W OPCPA system for high-repetition-rate experiments at the LCLS x-ray free-electron laser," Opt. Lett. **44**, 1257-1260 (2019)

20. S. Prinz, M. Haefner, C. Y. Teisset, R. Bessing, K. Michel, Y. Lee, X. T. Geng, S. Kim, D. E. Kim, T. Metzger, and M. Schultze1, "CEP-stable, sub-6 fs, 300-kHz OPCPA system with more than 15 W of average power", Opt. Express, **23**, 1388-1394 (2015)

21. C. Röcker, A. Loescher, F. Bienert, P. Villeval, D. Lupinski, D. Bauer, A. Killi, T. Graf, and M. A. Ahmed, "Ultrafast green thin-disk laser exceeding 1.4 kW of average power," Opt. Lett. **45**, 5522–5525 (2020)

22. J. Adamonis, R. Antipenkov, J. Kolenda, A. Michailovas, A.P. Piskarskas, A. Varanavičius, and A. Zaukevičius, "Formation of Flat-top picosecond pump pulses for OPCPA systems by cascade second harmonic generation", Lith. J. Phys. **52**, 193–202 (2012)

23. P. Mackonis, and A. M. Rodin, "OPCPA investigation with control over the temporal shape of 1.2 ps pump pulses", Opt. Express, **28**, 12020-12027 (2020)

24. J. Moses and S.-W. Huang, "Conformal profile theory for performance scaling of ultrabroadband optical parametric chirped pulse amplification," J. Opt. Soc. Am. B **28**, 812–831 (2011)

25. H. R. Telle, G. Steinmeyer, A. E. Dunlop, J. Stenger, D. H. Sutter, and U. Keller, "Carrier-envelope offset phase control: A novel concept for absolute optical frequency measurement and ultrashort pulse generation," Appl. Phys. B, **69**, 327-332 (1999)

26. J. Pupeikis, N. Bigler, S. Hrisafov, C. R. Phillips, U. Keller, "Programmable pulse shaping for time-gated amplifiers", Opt. Express, **27**, 175-184 (2019)

27. R. Eckardt and J. Reintjes, "Phase matching limitations of high efficiency second harmonic generation," IEEE J. Quantum Electron. **20**, 1178–1187 (1984)

28. J. Bromage, J. Rothhardt, S. Hädrich, C. Dorrer, C. Jocher, S. Demmler, J. Limpert, A. Tünnermann, and J. D. Zuegel, "Analysis and suppression of parasitic processes in noncollinear optical parametric amplifiers," Opt. Express **19**, 16797-16808 (2011)

29. A. Baltuška, M.S. Pshenichnikov, and D.A. Wiersma, "Second-harmonic generation frequency-resolved optical gating in the single-cycle regime", IEEE J. Quantum Electron. **35**, 459–478 (1999)